\documentclass[journal=ancac3,manuscript=article,layout=twocolumn]{achemso}
\usepackage{amsmath}
\usepackage{amssymb}

\usepackage{chemformula} 
\usepackage[T1]{fontenc} 
\usepackage{graphicx}
\usepackage{dcolumn}
\usepackage{ltxgrid}
\usepackage{color}
\usepackage[hidelinks]{hyperref}
\usepackage{bm}

\author{Joeri Opdam}
\affiliation[University of Tokyo]
{Research Center for Advanced Science and Technology, The University of Tokyo, 4-6-1 Komaba, Meguro-ku, Tokyo 153-8904, Japan}
\author{Michio Tateno}
\affiliation[University of Tokyo]
{Research Center for Advanced Science and Technology, The University of Tokyo, 4-6-1 Komaba, Meguro-ku, Tokyo 153-8904, Japan}
\alsoaffiliation[University of California]
{Materials Department, University of California, Santa Barbara, California 93106, USA}
\author{Hajime Tanaka}
\email{tanaka@iis.u-tokyo.ac.jp}
\affiliation[University of Tokyo]
{Research Center for Advanced Science and Technology, The University of Tokyo, 4-6-1 Komaba, Meguro-ku, Tokyo 153-8904, Japan}
\alsoaffiliation{Department of Fundamental Engineering, Institute of Industrial Science, The University of Tokyo, 4-6-1 Komaba, Meguro-ku, Tokyo 153-8505, Japan}

\title[]{Unraveling the impact of competing interactions on non-equilibrium colloidal gelation}

\keywords{Colloids, Gelation, Competing interactions, Microscopic structure, Bernal spiral, Depercolation\relax\footnotetext{This document is the unedited Author’s version of a Submitted Work that was subsequently accepted for publication in ACS Nano, copyright © 2025 American Chemical Society after peer review. To access the final edited and published work see https://doi.org/10.1021/acsnano.5c03244}}

\begin{document}

\onecolumngrid
\begin{abstract}
Competing interactions stabilize exotic mesoscopic structures, yet the microscopic mechanisms by which they influence non-equilibrium processes leading to disordered states remain largely unexplored, despite their critical role in self-assembly across a range of nanomaterials and biological systems. Here, we numerically investigate the structural evolution in charged colloidal model systems, where short-range attractions and long-range repulsions compete. We reveal that these two interaction scales drive sequential ordering within clusters, from tetrahedra motifs to linear aggregates with chiral order. This process disrupts early-stage percolated networks, resulting in reentrant behavior --- a dynamic transition from disordered cluster to network to chiral rigid cluster. On the other hand, the cluster-elastic network boundary in the final state is governed by isostatic percolation, which slows structural rearrangements, preserves branching points, and sustains a long-lived network. The resulting structure consist of rigid Bernal spiral-like branches connected through flexible branching points lacking order. These insights advance our microscopic understanding of out-of-equilibrium ordering driven by competing interactions, particularly phenomena like temporally delayed frustration reflecting different length scales of competing interactions.
The mechanisms identified here may play a crucial role in mesoscale self-organization across soft materials, from nanoparticle assemblies to biological gels and cytoskeletal networks. Understanding how competing interactions regulate structure and dynamics could guide the design of adaptive materials with tunable mechanical properties and offer new perspectives on biological processes such as cytoplasmic organization and cellular scaffolding.
\end{abstract}
\newpage
\twocolumngrid
\newpage

\section{Introduction}

Colloidal gels typically form through phase separation driven by interparticle attractions, resulting in a network of physically bound particles suspended in a fluid medium. Their unique combination of solid-like mechanical properties and fluid-like transport characteristics makes them invaluable across a range of applications, including cosmetics, inks, cement, pharmaceuticals, and energy storage devices~\cite{Piazza2011, Nelson2019, Smay2002, Ioannidou2016, Diba2018, Tagliaferri2021}. Beyond practical applications, colloidal gels serve as model systems for understanding protein gelation in complex biological environments, such as food science~\cite{Mezzenga2005, Nicolai2019, Patel2020} and intracellular organization~\cite{Berry2018, Boeynaems2018, Tanaka2022}.

A fundamental challenge in colloidal gel research is controlling their structure and functionality, which is essential both for both technological applications and understanding biological processes. A promising approach to achieving this control is the introduction of competing interactions, such as long-range repulsion and short-range attraction. These interactions introduce frustration --- the inability of a system to settle into a simple, energetically favorable configuration --- leading to a rich diversity of exotic states, complex structures, and unconventional dynamic behaviors ~\cite{seul1995domain,andelman2009modulated}. This phenomenon is widely observed across hard and soft condensed matter and biological systems and plays a central role in the emergence of novel material properties in nanostructured materials.

For example, in magnetic systems, competition between long-range repulsion (e.g., dipole-dipole interactions) and short-range exchange interactions profoundly influences phase transitions, magnetic ordering, and the emergence of topological phases~\cite{dagotto2005complexity,cheong2007multiferroics,balents2010spin}. In soft matter, similar mechanisms drive mesoscale pattern formation in polymers~\cite{muthukumar1997competing,harrison2000} and control colloidal self-assembly~\cite{tanaka2004nonergodic,li2016assembly}. In biological systems, they regulate protein folding pathways and influence charged membrane ordering~\cite{andelman1995electrostatic}.

Despite substantial progress in understanding how competing interactions govern equilibrium states, their impact on non-equilibrium ordering --- particularly in gelation dynamics --- remains poorly understood. Previous studies have explored these dynamics using coarse-grained models, such as phase-field simulations of microphase separation~\cite{ohta1986equilibrium,sagui1994kinetics}. However, such approaches, while effective in capturing large-scale structural evolution, often overlook crucial microscopic details, including topological ordering at the particle level. A deeper, particle-resolved approach is necessary to fully unravel the role of competing interactions in gelation.

In this study, we investigate colloidal systems with long-range Coulomb repulsion and short-range depletion attraction, providing a model framework for understanding how these interactions drive non-equilibrium gelation dynamics at the particle-level. Gaining insights into these mechanisms is not only of fundamental scientific interest but also crucial for designing materials that rely on controlled colloidal gelation.

To establish a baseline, we first consider the behavior of colloidal gelation driven by short-range attractions in the absence of competing interactions~\cite{Piazza1994,Poon1997,Poon2002,Zaccarelli2008,Lu2008,Royall2018a,Royall2021}. Such systems typically undergo phase separation, leading to aggregation in an unstable state. 
At low volume fractions, aggregates compact and solidify before percolating, forming stress-free gels~\cite{Tsurusawa2020}. Conversely, at higher volume fractions, early percolation induces mechanical stress, leading to viscoelastic phase separation~\cite{Tsurusawa2019, Tanaka2000, Tanaka2007}. Over time, network coarsening occurs via interparticle attraction and surface tension, following a self-similar power-law growth~\cite{Tateno2021,Tateno2022,tateno2025impact}. As the gel solidifies, this process slows, marking the onset of ageing~\cite{wang2024distinct}. Traditionally, this slowdown has been attributed to dynamic arrest, where local density approaches the glass transition volume fraction~\cite{Segre2001,Dawson2002,Kroy2004,Manley2005,Zaccarelli2007}. However, recent studies suggest that solidity instead arises from mechanical arrest~\cite{Hsiao2012,Tsurusawa2019,Zhang2019,Tateno2022,tsurusawa2023hierarchical}, as isostatic networks form with coordination numbers satisfying Maxwell's rigidity criterion~\cite{Maxwell1864}, similar to the emergence of solidity in glasses~\cite{tong2020emergent}.

This scenario changes significantly when long-range Coulomb interactions, introduced by colloid surface charges, come into play~\cite{bonn1999laponite}. These interactions introduce frustration, leading to the formation of complex structures, such as modulated phases and Bernal spirals~\cite{Bernal1964,Campbell2005,Sciortino2005}, and significantly altering gelation dynamics~\cite{Sedgwick2004,Campbell2005,Sciortino2005,Zaccarelli2007,Cao2010,Valadez-Perez2013,Liu2019,Ruiz-Franco2021}. Notably, electrostatic repulsions suppress the formation of dense structures, yielding gels with smaller pores, thinner branches, and altered particle arrangements.

At sufficiently strong repulsion, equilibrium cluster phases with well-defined sizes and shapes emerge below the gelation threshold~\cite{Groenewold2001,Stradner2004,Sciortino2004,Mossa2004,Liu2011,Godfrin2014,Bollinger2016a,Bollinger2016b}. As these clusters grow, their dynamics slow, leading to kinetically arrested states such as cluster fluids or Wigner glasses~\cite{Toledano2009,Klix2010,Zhang2012,Mani2014,Ruiz-Franco2021}.
Gelation in such systems is often associated with a geometrical percolation threshold, wherein compact clusters connect to form a spanning network~\cite{Zaccarelli2007,Ruiz-Franco2021}.
While prior research has characterized the structure of charged colloidal gels in their ageing regime, the influence of competing interactions on their kinetic evolution --- both locally and globally --- remains largely unexplored.

Here, we investigate how long-range repulsions reshape the kinetic pathway of gelation in colloidal systems undergoing phase separation due to short-range attractions at the microscopic level. Our findings uncover a counterintuitive process at intermediate volume fractions, where gelation follows an inverse pathway: rather than compact clusters percolating into a gel, an initially percolated network forms through random particle connections but subsequently fragments into distinct clusters over time. 
This unconventional behavior arises from sequential ordering processes driven by interactions at different length scales. In the early stages, short-range attractions dominate, leading to network formation. At later stages, long-range repulsions take over, restructuring and fragmenting the network. In gel-forming regimes, this interplay also gives rise to distinctive structural characteristics: gel branches develop rigid linear Bernal spiral-like structures with pronounced chirality, while branching points remain disordered and flexible. These findings not only enhance our understanding of gelation under competing interactions but also offer new strategies for designing colloidal materials with tunable mechanical properties.

\section{Results and discussion}
\begin{figure}[t!]
\includegraphics[width=.49\textwidth]{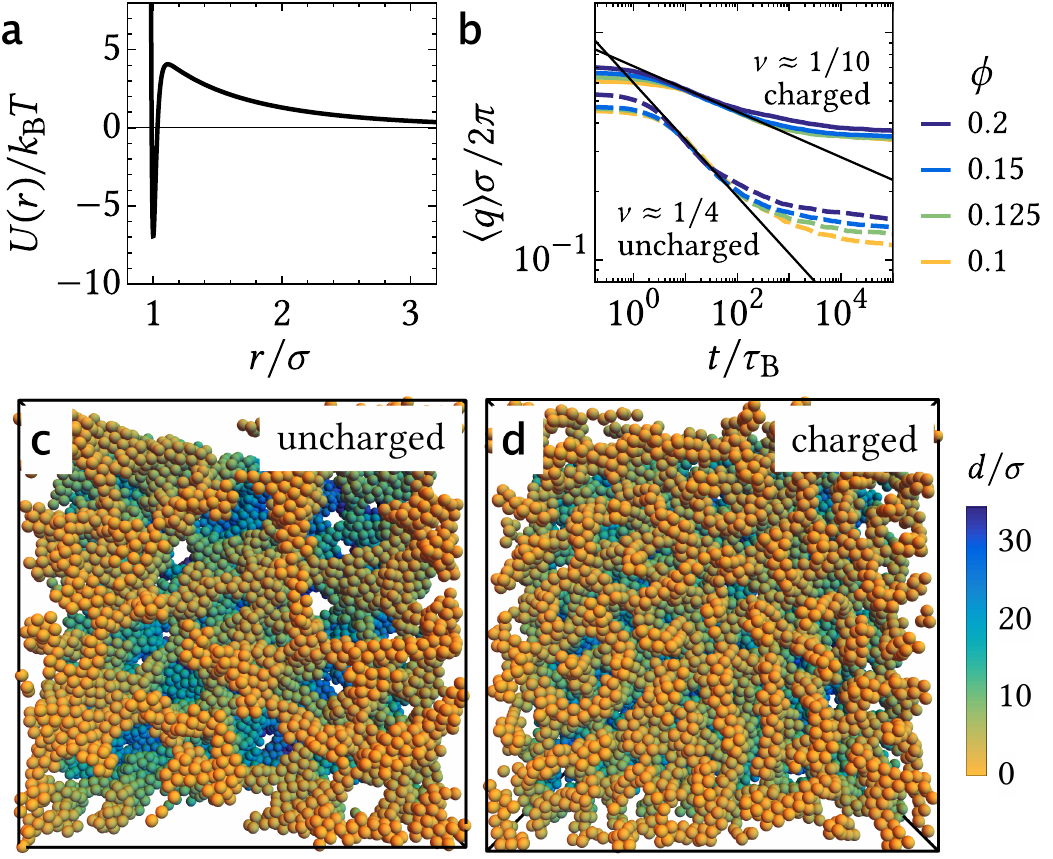}
\caption{{\bf Comparison of global structure in charged and uncharged colloidal systems.}
\textbf{a} Interaction potential $U(r)$ as a function of the distance $r/\sigma$ for the studied charged systems. \textbf{b} Evolution of the characteristic wavenumber $\langle q\rangle$ over time $t$, normalized by the Brownian time scale $\tau_\text{B}$ for charged (solid curves) and uncharged systems ($U(0)_\text{r}=0$, dashed curves). \textbf{c}, \textbf{d} Snapshots of gels formed after $t=10^5\tau_\text{B}$ in uncharged and charged systems with volume fraction $\phi=0.2$. Colour indicates the depth $d$ within the simulation box.}
\label{f1}
\end{figure}

We investigate the structural evolution in charged colloidal systems using Langevin Dynamics simulations. The intercolloidal interactions are modeled using a combination of a short-range Morse potential and a long-range Yukawa interaction:
\begin{multline*}
    U(r)=U(0)_\text{r}\frac{\sigma}{r}\exp\left[-\kappa\sigma\left(r/\sigma-1\right)\right]   \\
    +U(0)_\text{a}(\exp\left[-2\alpha\left(r/\sigma-1\right)\right]-\\2\exp\left[-\alpha\left(r/\sigma-1\right)\right]),
\end{multline*}
with $r$ the interparticle distance. These parameters mimic a dispersion of charged colloidal particles interacting via a depletion attraction induced by non-adsorbing polymers ~\cite{Lekkerkerker2024}. The range of the attraction is roughly $0.1\sigma$, corresponding to $\alpha = 45$, and the Debye length is $\kappa^{-1}=1.5\sigma$, with $\sigma$ representing the diameter of the colloidal particles. The contact potentials of the depletion attraction and electrostatic repulsion are chosen as $U(0)_\text{a} = -12 \, k_\text{B}T$ and $U(0)_\text{r} = 5 \, k_\text{B}T$, respectively. The interaction potential is depicted in Fig.~\ref{f1}a. For a selection of results, a comparison is made with an uncharged system where $U(0)_\text{r}=0$. 

The chosen interaction parameters fall within the experimental range studied previously~\cite{Liu2019}. The Debye length and contact potential values are situated between those reported in experimental systems studied by Zhang \textit{et al.}~\cite{Zhang2012} and Campbell \textit{et al.}~\cite{Campbell2005}. Experimental observations in these systems have indicated the formation of elongated and Bernal spiral-like structures. The Bernal spiral is a chiral, linear cluster with a triple-helix geometry, consisting of at least four tetrahedral subunits connected face-to-face in a linear arrangement~\cite{Bernal1964}. Previous studies~\cite{Mossa2004} have shown that for certain combinations of short-range attractions and long-range repulsions, the ground-state configuration of clusters is the Bernal spiral, a configuration prominently observed in the systems studied here.

\subsection*{Global structural evolution}

First, we discuss the development of global structures observed in the studied systems. The characteristic spacing between clusters or network branches is characterized by the characteristic wavenumber $\langle q\rangle$, defined as the first moment of the structure factor $S(q)$: $\langle q\rangle = \int qS(q)\text{d}q / \int S(q)\text{d}q$. Figure~\ref{f1}b shows the temporal evolution of $\langle q\rangle$ for various volume fractions $\phi$, corresponding to either a gel state ($\phi=0.15$ and $\phi=0.2$) or a cluster state ($\phi=0.1$ and $\phi=0.125$). We define the gel state here by the presence of a percolated network in the final investigated timestep $t=10^5\tau_\text{B}$. The characteristic wavenumber evolves in a similar manner over time as in purely attractive colloidal systems shown in Fig.~\ref{f1}b for comparison: an initial constant stage, a coarsening stage with a power-law decrease ($\langle q\rangle \sim t^{-\nu}$), and a final aging stage with a slow logarithmic decay. However, the exponent of the power-law growth $\nu$ is much smaller than that of the purely attractive colloidal systems. This is due to interparticle electrostatic repulsions that slow down particle aggregation and cluster/network compaction. Furthermore, the thin cluster/network structure of charged particles compared to that of an uncharged gel, as demonstrated in Fig.~\ref{f1}c and d, results in a smaller pore size and thus a higher value of $\langle q\rangle$ in the aging regime.

\begin{figure}[t!]
\includegraphics[width=.49\textwidth]{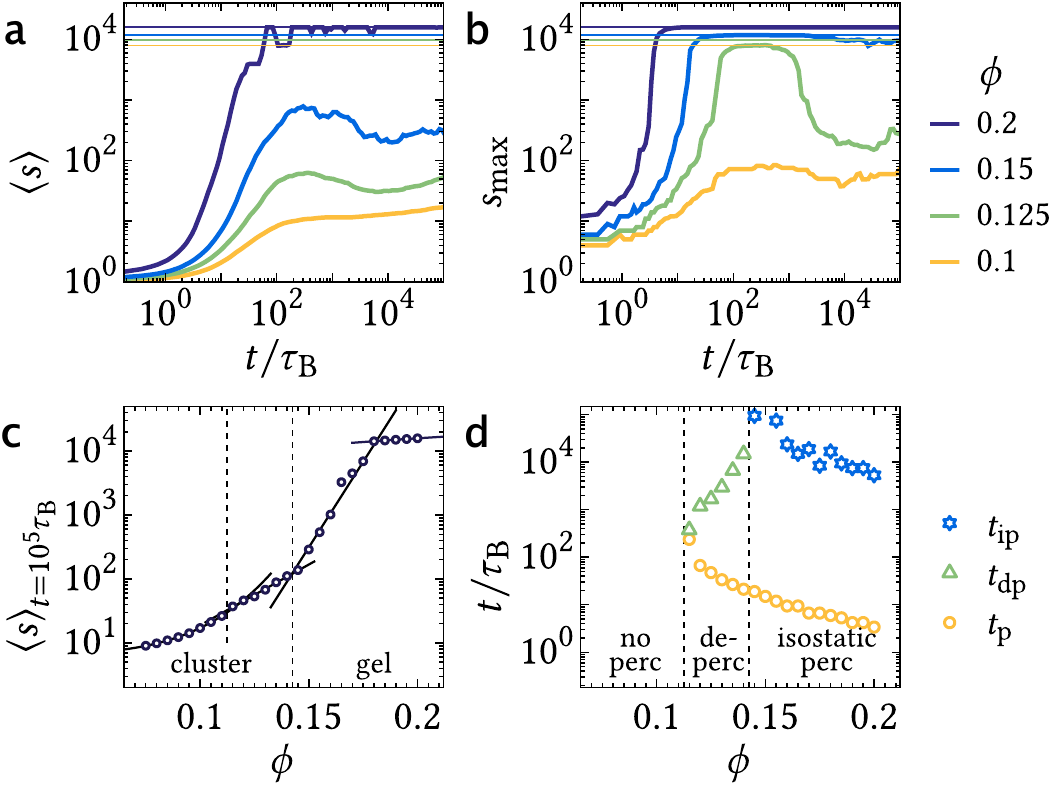}
\caption{{\bf Characterization of cluster size during structure evolution.}
\textbf{a}, \textbf{b} Evolution of the average cluster size (in terms of the number of particles), $\langle s\rangle$, and maximum cluster size, $s_\text{max}$, over time $t$, normalized by the Brownian time scale $\tau_\text{B}$. Horizontal lines indicate the number of particles, $N_\text{p}$, in the systems for the investigated volume fractions. \textbf{c} Average cluster size after $t=10^5\tau_\text{B}$ as a function of the colloid volume fraction $\phi$. Solid curves highlight different scaling regions and vertical dashed lines indicate the boundaries between observed structure evolution pathways. \textbf{d} Percolation ($t_\text{p}$), depercolation ($t_\text{dp}$), and isostatic percolation ($t_\text{ip}$) times as a function of $\phi$.} 
\label{f2}
\end{figure}

Previous studies have shown minimal differences in the radial distribution function~\cite{Klix2010} and structure factor~\cite{Ruiz-Franco2021,Sciortino2005} between a cluster state at low volume fractions and a gel state at high volume fractions in colloidal systems with competing interactions. These structural similarities between the cluster and gel states are also evident in Fig.~\ref{f1}b, which additionally demonstrates that the evolution of the global network structure over time is similar for both states. Prior research~\cite{Tateno2021,tateno2025impact} has indicated that, in the phase separation of uncharged colloidal systems, hydrodynamic interactions significantly influence the coarsening dynamics and are essential for reproducing the characteristic scaling law, $\langle q\rangle\sim t^{-1/2}$, observed in experiments. In Sec.~1 of the Supporting Information, we briefly compare the results shown in Fig.~\ref{f1}b (and Fig.~\ref{f2}a and b) with those from Fluid Particle Dynamics simulations~\cite{Tanaka2000} that include many-body hydrodynamic interactions. For the charged system studied here, our preliminary results indicate that incorporating hydrodynamic interactions does not significantly alter the power-law coarsening behavior. 
This may be attributed to interparticle repulsions diminishing the impact of squeezing flow, which is influenced by the solvent's incompressibility, on colloidal aggregation. A more detailed investigation into the interplay between hydrodynamic and electrostatic interactions and its effect on colloidal gelation is currently ongoing.

The time evolution of the charged systems in terms of average cluster size ($\langle s\rangle$) and maximum cluster size ($s_\text{max}$), where $s$ denotes the number of particles within a cluster, is depicted in Fig.~\ref{f2}a and b. For each volume fraction, both parameters increase rapidly in the early stage. However, in the late stage, the systems exhibit qualitative differences depending on the volume fraction $\phi$. 

\begin{figure*}[t!]
\includegraphics[width=.99\textwidth]{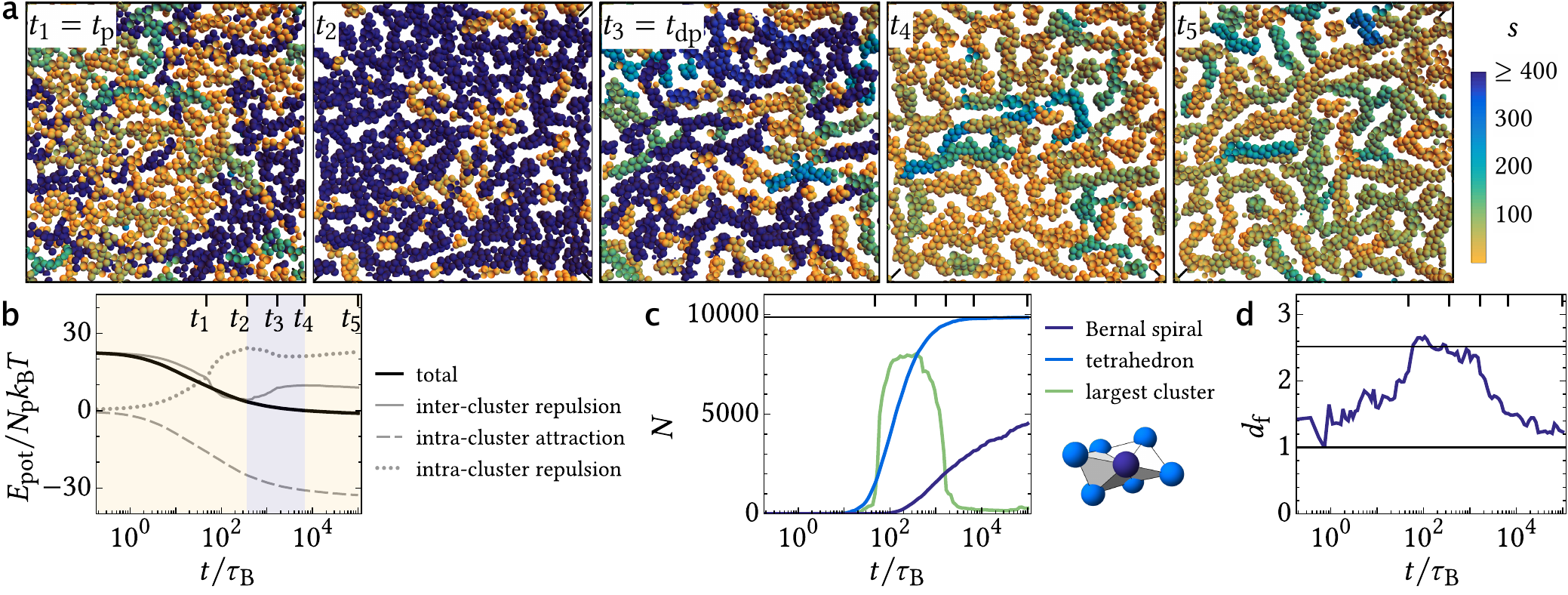}
\caption{{\bf Structural evolution accompanying a percolation-depercolation transition.}
\textbf{a} Snapshots of the simulated system with a colloid volume fraction of $\phi=0.125$ for the time steps described in the main text. For visualization, slices with $0.2$ times the depth of the simulation box are shown. \textbf{b} Time evolution of the potential energy $E_\text{pot}$ of the system. Background colours distinguish regions where the average cluster size is increasing (yellow) and decreasing (purple). \textbf{c} Time evolution of the number of particles $N$ that are part of the largest cluster, part of at least one tetrahedron, and part of exactly four tetrahedra connected face-to-face in a Bernal spiral configuration, as shown in the schematic picture below the legend. \textbf{d} Time evolution of the fractal dimension $d_{\rm f}$. The horizontal lines correspond to $d_\text{f}=2.52$ and $d_\text{f}=1$, representing random percolation and linear clusters, respectively.}
\label{f3}
\end{figure*}

We identify three distinct structure evolution pathways. For $\phi=0.1$, a space-spanning percolated network does not form. Instead, particles organize into relatively small clusters that remain disconnected over time. The average cluster size gradually increases in the late stage, indicating that cluster growth slows down due to long-range repulsive interactions between clusters. This behavior contrasts with the formation of equilibrium clusters, where a plateau value of $\langle s \rangle$ would be expected in the late stage.
At intermediate concentrations, an increase in the average cluster size is also observed in the late stage. However, this increase is preceded by a stage where the average and maximum cluster sizes significantly decrease. For the volume fraction of $\phi=0.125$, the number of particles in the largest cluster decreases by a factor of more than 10, resulting in a complete breakup of the previously formed network structure. We refer to this remarkable structure evolution as ``depercolation,'' which is discussed in more detail in the context of Fig.~\ref{f3}. 
For $\phi=0.15$ and $\phi=0.2$, depercolation does not occur, and the network remains intact throughout the late stage. These three characteristic temporal structure evolutions for different volume fractions can be seen in Supporting Movies 1--4.

Figure~\ref{f2}c shows the average cluster size after $10^5\tau_\text{B}$ for a range of volume fractions. At high volume fractions ($\phi \geq 0.18$), all particles are part of the network, and the average cluster size is equal to the total number of particles in the system. For lower volume fractions, three regions with different scaling of the average cluster size as a function of $\phi$ can be distinguished. The boundaries of these concentration regions roughly correspond to the boundaries between the three different structure evolution pathways described in the previous paragraph. This is shown in Fig.~\ref{f2}d, where the percolation, depercolation, and isostatic percolation times for different volume fractions are illustrated. The scaling behavior of cluster size as a function of volume fraction thus depends on whether the cluster state undergoes an intermediate percolation stage. In the depercolation regime, the depercolation time $t_\text{dp}$ exponentially increases with $\phi$ until a certain volume fraction, beyond which depercolation is no longer observed, and the network structure remains intact. This volume fraction coincides with the lowest volume fraction for which isostatic percolation, corresponding to a space-spanning network consisting of particles with at least six nearest neighbours, is observed.

The results shown in Fig.~\ref{f2} indicate that percolation does not mark the boundary between the cluster and gel states but rather suggests the existence of a crossover region characterized by relatively large clusters that undergo an intermediate percolation and depercolation stage. This implies that gelation in charged systems is not explained by a cluster state crossing the percolation boundary, but instead by mechanically arrested microphase separation. A similar gelation mechanism was previously suggested by Charbonneau and Reichman~\cite{Charbonneau2007}, who demonstrated that, for their systems, a cluster state is more energetically favorable than the network structures observed in Molecular Dynamics simulations. To our knowledge, depercolation has not been observed experimentally; however, Klix \textit{et al.}~\cite{Klix2010} reported the breakup of large clusters into smaller ones in the aging regime of a cluster state of colloids with competing interactions. The authors also measured the variance in cluster size, $\langle s^2 \rangle$, for different concentrations and reported a logarithmic scaling around the cluster/gel transition similar to Fig.~\ref{f2}c. However, the number of data points in their study is too low to distinguish between different scaling regimes and make a detailed comparison with our results.

\subsection*{Mechanism of a percolation-depercolation transition}

Now, we focus on the percolation-depercolation transition observed at intermediate volume fractions. Figure~\ref{f3} provides a closer look at the structural evolution of a colloidal system with $\phi=0.125$, where depercolation occurs. Figure~\ref{f3}a shows snapshots of the system at various time points: at percolation time ($t_\text{1}=t_\text{p}$), when the average cluster size $\langle s \rangle$ starts to decrease ($t_\text{2}$), at depercolation time ($t_\text{3}=t_\text{dp}$), when the average cluster size starts increasing again ($t_\text{4}$), and at the final investigated time regime ($t_\text{5}=10^5\tau_\text{B}$). Cluster size distributions and the corresponding average potential energy per particle for clusters of different sizes at these times are shown in Supporting Fig.~S2. 

In addition to the drastic changes in the mean cluster size over time, a notable increase in the local structure is observed. The development of the system's potential energy over time is shown in Fig.~\ref{f3}b. The potential energy trend mirrors that of the characteristic structure factor in Fig.~\ref{f1}b. Also depicted are the separate contributions of attractive and inter/intra-cluster repulsive interactions. The attractive part of the potential energy follows the same path as the total potential energy, indicating that the average coordination number increases over time, even during the depercolation stage. The number of intra-cluster repulsive interactions decreases during this stage, suggesting a structural reorganization that minimizes long-range repulsive interactions while still increasing the number of attractive interactions.

The structural reorganization is further highlighted in Fig.~\ref{f3}c, which shows the number of particles that are part of the largest cluster, part of a tetrahedron, and part of exactly four face-sharing tetrahedra in a Bernal spiral morphology. In the percolation stage, there are relatively few tetrahedral structures present in the system, and thus, almost no Bernal spirals are formed. This shows that network formation does not occur through the percolation of clusters close to their ground state energy. During network formation, tetrahedral structures begin to develop. Only in later stages, roughly from the time when the network starts breaking up, do the tetrahedra start to form long-range order in the shape of Bernal spiral-like structures. This demonstrates that there are two distinct time scales in the structural evolution: an attraction-driven early stage of percolation where both attractive and repulsive interactions rapidly increase, and a repulsion-driven late stage where long-range order is established and the contribution of electrostatic repulsions is minimized.

The time evolution of the fractal dimension $d_{\rm f}$ is shown in Fig.~\ref{f3}d.
Initially, during percolation, the fractal dimension increases, reaching $d_{\rm f} \approx 2.52$, indicative of random percolation. After the depercolation time, $d_{\rm f}$ decreases to around $d_{\rm f} \approx 1$, suggesting that the network structure breaks up around branching points, leading to the formation of highly linear clusters. In the final stage of structure development (after $t_\text{4}$), the average cluster size starts to increase again, as shown in Fig.~\ref{f2}a. However, no percolated network forms during this second growth phase; instead, individual clusters aggregate linearly, as evidenced by the lack of increase in the fractal dimension between $t_\text{4}$ and $t_\text{5}$.

\subsection*{Local structural evolution}

Next, we explore the development of unique local structures, arising from competing interactions. Figure~\ref{f3} illustrates the evolution toward linear clusters with a Bernal spiral-like structure. This is evident from the large number of particles in the late stage that exhibit a local environment resembling a Bernal spiral. Distributions of the bond order parameters $q_6$, $q_4$, $\hat{w}_4$, and $\hat{w}_6$ of the studied systems also exhibit distinct peaks associated with a Bernal spiral structure in both the cluster and gel states, as shown in Supporting Fig.~S3. 

\begin{figure}[t!]
\includegraphics[width=.49\textwidth]{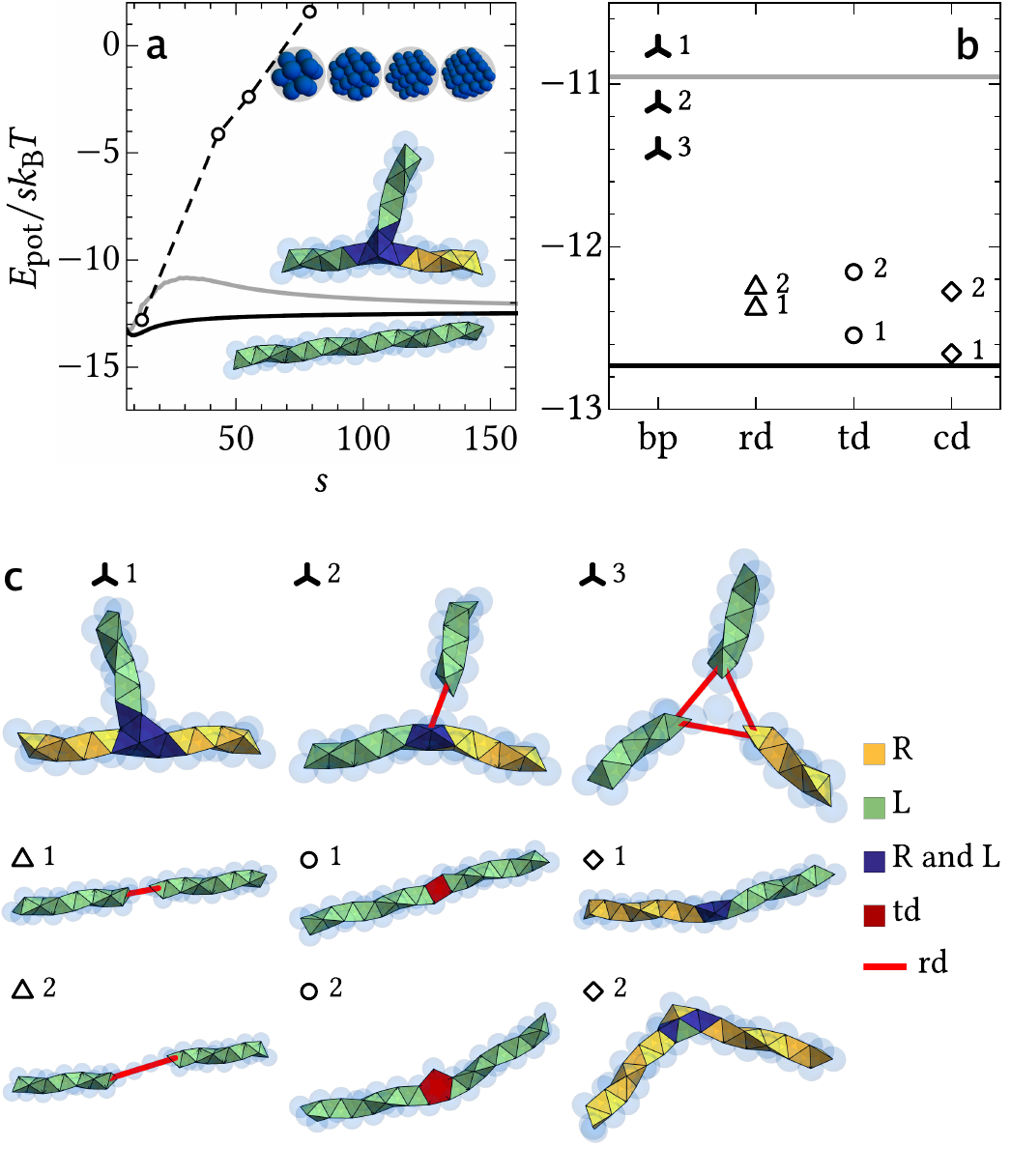}
\caption{{\bf Overview of defects leading to deviations from Bernal spiral structure.}
\textbf{a} Potential energy per particle for isolated clusters with spherical (symbols, dashed curve as a guide to the eyes), Bernal spiral (black curve) and T-shaped Bernal spiral (gray curve) morphologies, as a function of cluster size $s$. \textbf{b} Potential energy per particle for clusters shown in panel \textbf{c} that contain branching points (bp), rigidity defects (rd), tetrahedral defects (td), or chiral defects (cd). Horizontal lines represent the potential energy of linear (black) and T-shaped Bernal (gray) spirals with the same number of particles ($s=40$). \textbf{c} Examples of clusters exhibiting these defects. Tetrahedra are part of a right-handed Bernal spiral (yellow), a left-handed Bernal spiral (green), or, due to a chiral defect, both a left-handed and right-handed Bernal spiral (purple).}
\label{f4}
\end{figure}

Fig.~\ref{f4}a presents the potential energy of isolated clusters (considering only intra-cluster interactions) as a function of cluster size for three different cluster morphologies: face-centerd cubic (spherical), Bernal spiral, and T-shaped Bernal spiral. The Bernal spiral is significantly more favorable than the spherical geometry due to the reduction of electrostatic repulsions in the linear morphology. The T-shaped Bernal spiral, representing a rigid branching point between Bernal spirals, is less favorable compared to the linear cluster because particles at branching points experience relatively stronger repulsive interactions than those in a linear configuration. This observation aligns with the fact that a percolated network breaks into highly linear clusters during depercolation. While the potential energy curve for the linear Bernal spiral suggests an optimal cluster size of about ten particles, this finding is relevant primarily in the dilute limit and does not fully account for the inter-cluster interactions significant in the systems studied here.

We will characterize local structures by identifying defects that cause deviations from the favored Bernal spiral structure. To define these defects, we introduce a structural building block termed a ``rigid cluster (rc).'' A rigid cluster consists exclusively of tetrahedral, pyramid, and pentagonal pyramid subunits that share at least three common particles with neighbouring subunits. These clusters are termed ``rigid'' because each particle is in a locally favored configuration, and any translational movement that does not involve every subunit of the cluster necessitates breaking inter-particle bonds.

A ``rigidity defect (rd)'' is defined as a bond between two rigid clusters that are connected through one or two common particles, particles considered nearest neighbours, or a path consisting of neighbouring particles not part of a rigid cluster. Such a rigidity defect permits independent motion of the two rigid clusters within the larger structure without breaking inter-particle bonds.

Within rigid clusters, two types of defects can deviate from the Bernal spiral structure. The first is a ``tetrahedral defect (td),'' which occurs when pyramid or pentagonal pyramid structures replace tetrahedra (note that a pentagonal bipyramid built from five tetrahedra is not regarded as a tetrahedral defect). The second is a ``chiral defect (cd),'' which arises when tetrahedra are connected face-to-face but experience a change in the dihedral angle between their centers, resulting in a shift in chirality. Examples of clusters exhibiting these defects are shown in Fig.~\ref{f4}c.

The potential energy of these clusters, compared to linear and T-shaped Bernal spirals with the same number of particles, is illustrated in Fig.~\ref{f4}b. Also shown are different types of branching points (bp) between rigid clusters, which can be either fully rigid or have one or more rigidity defects. Unlike linear clusters, rigidity defects at branching points lead to lower potential energy. 
This reduction is partly due to fewer particles at the branching points experiencing relatively strong repulsive interactions and partly because the defects allow the branches to move independently, optimizing their spacing and minimizing repulsive interactions.

\begin{figure}[t!]
\includegraphics[width=.49\textwidth]{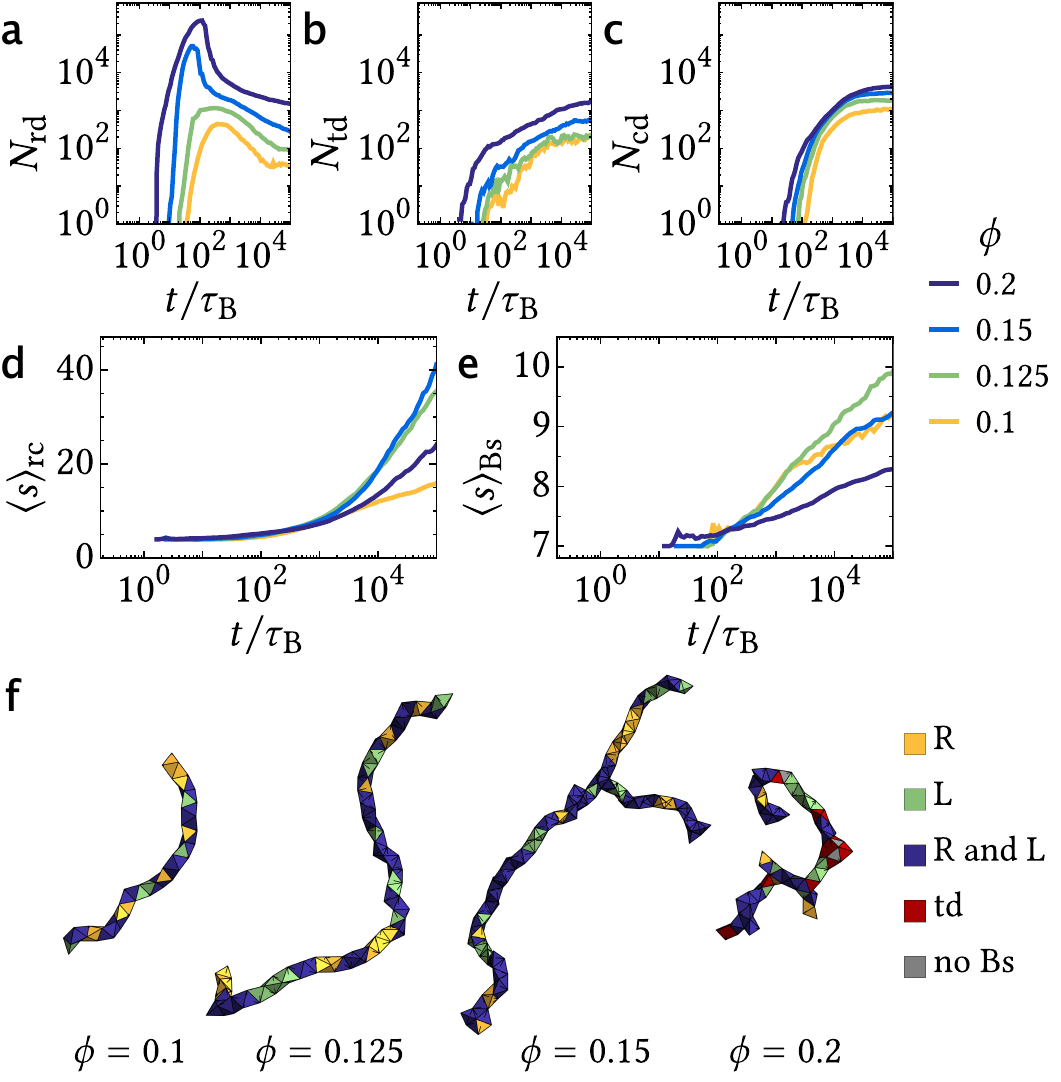}
\caption{{\bf Local structural evolution over time.}
\textbf{a}--\textbf{c} Time evolution of the number of rigidity defects, tetrahedral defects, and chiral defects, as defined in the main text. \textbf{d}, \textbf{e} Time evolution of the average size (in terms of the number of particles) of rigid clusters, $\langle s\rangle_\text{rc}$, and unique Bernal spirals, $\langle s\rangle_\text{Bs}$. \textbf{f} Local structure of the largest rigid cluster present in the simulated systems after $t=10^5\tau_\text{B}$. Colours follow the same scheme as in Fig.~\ref{f4}c, with tetrahedra not part of a Bernal spiral in gray.}
\label{f5}
\end{figure}

Fig.~\ref{f5}a--c illustrate the evolution of rigidity defects ($N_\text{rd}$), tetrahedral defects ($N_\text{td}$), and chiral defects ($N_\text{cd}$) over time. Initially, small rigid structures form randomly within percolated clusters, leading to a sharp peak in rigidity defects. As time progresses, the number of rigidity defects decreases due to either the breakup of clusters into disconnected rigid clusters or due to compaction and merging of rigid clusters. As the volume fraction increases, rigidity defects become significantly more prevalent. This is due to the rise in the number of branching points, which hinders compaction since rigid branching points are energetically unfavorable. Additionally, the breakup of rigidity defects around branching points and the subsequent merging of rigid clusters into a linear arrangement is significantly slowed down in the gel state. This slowdown is due to isostatic percolation and the higher energy barriers associated with the restructuring process at larger volume fractions.

The number of tetrahedral defects also increases significantly with volume fraction, especially in the gel state (Fig.~\ref{f5}b), indicating that percolation increases the probability for particles to be trapped in unfavorable non-tetrahedral structures. However, tetrahedral defects are less prevalent compared to chiral defects (compare Fig.~\ref{f5}b and c), which are less influenced by volume fraction and are already common in the cluster state. Chiral defects, which have a relatively low energetic cost, form easily when an additional particle attaches to an incorrect face of a pre-existing Bernal spiral. Furthermore, short Bernal spirals formed in the early stage consist of an equal mix of L- and R-handed spirals, and the merging of spirals of opposite chirality inevitably leads to chiral defects.

Fig.~\ref{f5}d and e show the average sizes of rigid clusters and unique Bernal spirals over time. A Bernal spiral is considered unique if it does not completely overlap with a larger Bernal spiral. Both the average sizes of rigid clusters and Bernal spirals increase throughout the depercolation stage, indicating that depercolation does not involve the breakup of rigid structures. The continuous growth in their average sizes, without plateauing in the late stage, suggests ongoing optimization of local structure in both the cluster and gel states during the aging regime.

For the cluster state ($\phi=0.1$ and $0.125$), the average size of rigid clusters is similar to the average size of regular clusters in the final stages, with rigid clusters becoming larger as the volume fraction increases. In the gel state ($\phi=0.15$ and $0.2$), however, the average size of rigid clusters is much smaller than that of regular clusters and decreases as the gel network becomes denser. This further implies that rigid, locally ordered structures do not persist around network branching points, which become more prevalent at higher volume fractions. 

The rigid structures obtained in both the cluster and gel states resemble Bernal spirals, but the average size of Bernal spirals is relatively small compared to that of the rigid clusters. Tetrahedral and chiral defects result in short-lived chirality within the rigid clusters, causing deviations from a purely linear structure characteristic of Bernal spirals. Figure~\ref{f5}f illustrates the largest rigid clusters present in the studied systems after $10^5\tau_\text{B}$. These rigid clusters exhibit a similar quasi-linear structure for the four different volume fractions, akin to the Bernal spiral-like structures seen in experimental systems~\cite{Campbell2005,Zhang2012}. The defects observed persist over long time scales and approach a plateau, inhibiting the formation of ordered modulated structures, specifically, columnar arrangements of aligned Bernal spirals, which would likely be the most energetically favorable state~\cite{Candia2006}.

\begin{figure}[t!]
\includegraphics[width=.49\textwidth]{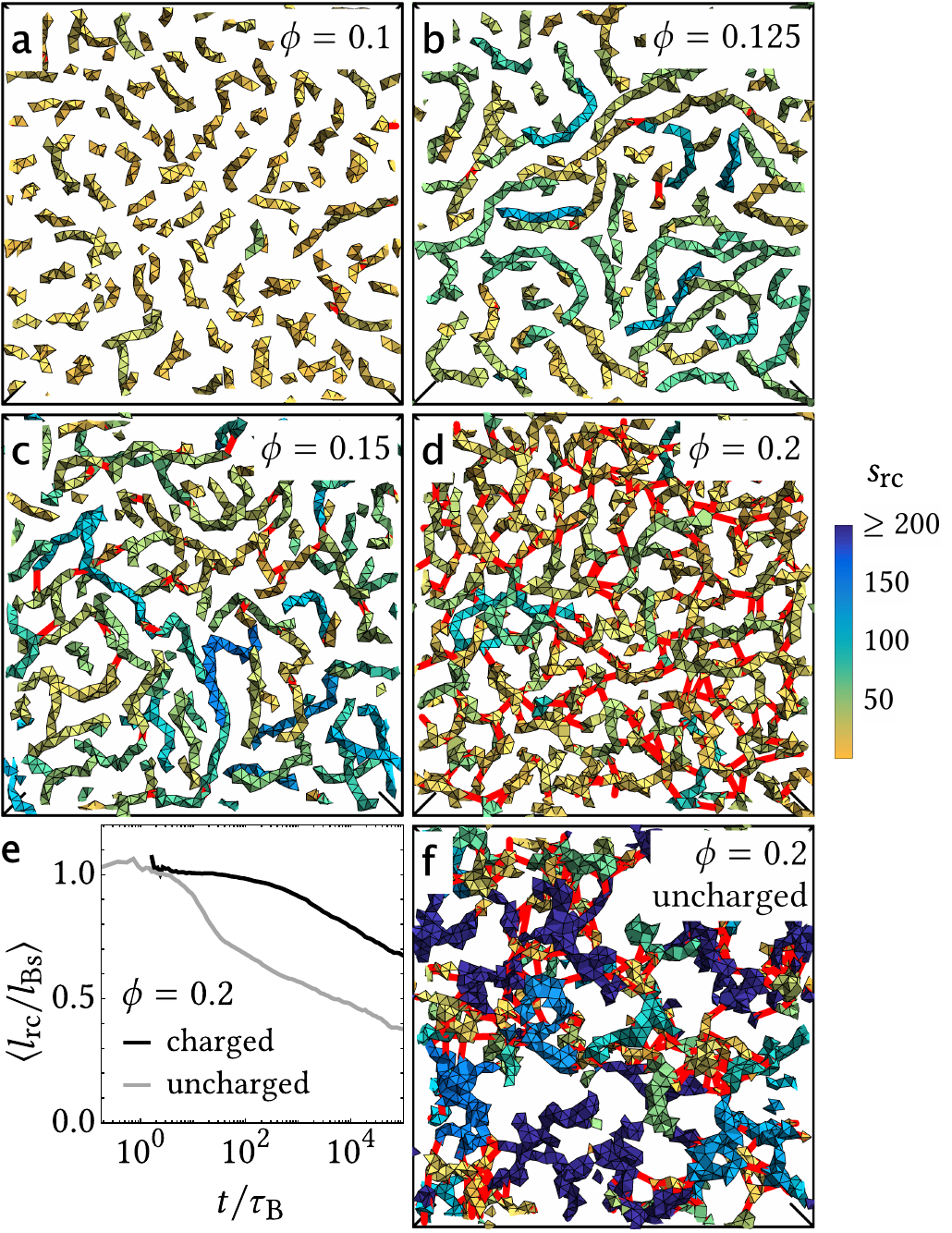}
\caption{{\bf Rigid structures in charged and uncharged colloidal systems.}
\textbf{a}--\textbf{d} Snapshots of the simulated charged systems after $t=10^5\tau_\text{B}$, showing tetrahedral, pyramid and pentagonal pyramid structures, coloured by rigid cluster size. Rigidity defects are indicated as thick red lines. For visualization, slices with $0.2$ times the depth of the simulation box are displayed. \textbf{e} Time evolution of the average relative linearity of rigid clusters $\langle l_\text{rc}/l_\text{Bs} \rangle$. \textbf{f} Same as \textbf{d} but for the uncharged system with $\phi=0.2$.}
\label{f6}
\end{figure}

Although an isostatic network is present in the gels, where each particle has an average of six or more neighbours, the average size of rigid clusters remains relatively small. The structural details of the studied systems are further highlighted in Fig.~\ref{f6}a--d which shows snapshots of the systems after $10^5 \tau_\text{B}$, with rigid clusters distinguished by colour and rigidity defects marked by thick red lines. The primary differences between the cluster and gel states are not in the rigid structures themselves but the number of flexible bonds connecting these rigid structures. Charged colloidal gels thus exhibit distinct structural characteristics: the branches are linear, rigid and locally ordered, while the branching points, which give rise to the network structure, are flexible and lack local order. 

Rigidity defects are associated with mechanical stress, which arises from a gelation pathway where percolation occurs before the compaction of clusters~\cite{Tsurusawa2020}. Such defects are also present in uncharged colloidal gels with volume fractions $\phi>0.1$, which follow a similar pathway. However, in uncharged colloidal gels, there is no preference for the linear growth of rigid clusters, and thus, no potential energy cost associated with rigid branching points. Consequently, rigid structures in uncharged colloidal gels are generally larger and less linear. This is illustrated in Fig.~\ref{f6}e and f, which show the relative linearity of rigid clusters over time for both uncharged and charged systems with $\phi=0.2$, and a snapshot of the uncharged system after $10^5\tau_\text{B}$. Relative linearity is defined here as an average of the ratio of the maximum particle-particle distance within a rigid cluster ($l_\text{rc}$) to the maximum particle-particle distance in a perfect Bernal spiral with an equal number of particles ($l_\text{Bs}$). For perfectly linear Bernal spirals, this ratio would be $l_\text{rc}/l_\text{Bs}=1$, and deviations from linearity in rigid clusters result in lower values. Figure~\ref{f6}e shows that rigid clusters in the uncharged system are less linear, which can also be seen by comparing Fig.~\ref{f6}d and f. Furthermore, a comparison of these two figures reveals that the size of rigid clusters is significantly larger in the uncharged system.

\section{Conclusions}

In summary, our study of the structure evolution in a charged colloidal system reveals that gelation onset, as a function of volume fraction, is marked by an isostatic percolation boundary. Percolation occurs before the formation of locally ordered structures. Below the isostatic percolation threshold, networks break up during compaction around energetically unfavorable branching points, resulting in linear clusters. This depercolation reflects delayed frustration arising from the interplay of competing interactions with different spatial ranges. Subsequently, a slow growth phase ensues, during which clusters aggregate in linear arrangements. Above this threshold, the restructuring process slows down significantly, preserving branching points and leading to a long-lived gel state. In the aging regime, both cluster and gel states show distinct linear arrangements of tetrahedral structures similar to Bernal spirals. However, chiral and tetrahedral defects lead to a short persistence of chirality, causing deviations in linear shape and preventing the formation of modulated ordered structures composed of long, linear Bernal spirals.

For the interaction potential examined here, local order does not persist around the branching points of the gel. As a result, the size of locally ordered regions decreases with increasing colloid concentration in the gel state. The results reveal an intriguing gel structure composed of rigid branches (strands) and relatively flexible branching points. This structure motivates further investigation into the correlation between structure and mechanical properties in charged colloidal gels. The generality of the observed gel structure and structural dynamics in charged colloidal systems with different interaction potentials and preferred cluster morphologies remains an important topic for further study.

Our work  highlights the importance of microscopic approaches. While a coarse-grained phase-field model with competing interactions may reproduce global ordering, it cannot capture the particle-level local ordering (e.g., chiral ordering) that is crucial to understanding the system's detailed behavior.

The competition between short-range attractions and long-range repulsions is known to drive the formation of various exotic equilibrium structures, such as micro-phase separation~\cite{seul1995domain}. Our study demonstrates how this interplay profoundly influences non-equilibrium processes, particularly gelation and cluster formation, from a microstructural perspective. Striking examples are the percolation-depercolation transition and flexible branching points, resulting from a sequential ordering process dominated initially by short-range attractions and subsequently by long-range repulsions. Additionally, we emphasize that gels formed under competing interactions have much thinner linear strands with distinct local and chiral order compared to uncharged gels. 

Introducing competing interactions in colloidal gels offers a powerful means to tailor their local and global structures, even at the same volume fraction, thereby optimizing both fluid transport and mechanical properties. 
Fine-tuning the strength and range of these interactions opens up exciting possibilities for enhancing the functionalities and performance of colloidal gels across various applications, paving the way for innovative advancements in material science. 
Beyond material engineering, our findings also shed light on the mechanisms underlying the formation of biological condensates in cells~\cite{Berry2018,Boeynaems2018,Tanaka2022}. We thus anticipate that the sequential ordering dynamics --- the delay in frustration effects --- arising from competition between different interaction scales, as observed in this study, could have broad relevance across a wide range of systems, including hard, soft, and biological materials.

\section{Methods}
\subsection*{Langevin Dynamics simulations} 
Langevin Dynamics (LD) simulations are performed using LAMMPS~\cite{LAMMPS} with an NVT ensemble and Langevin thermostat. The time step of the simulations is $\Delta t=4.8\cdot10^{-4}\tau_\text{B}$, with $\tau_\text{B} = \sigma/24D$, where $D=\frac{k_\text{B}T}{3\pi\sigma\eta_\text{s}}$ is the translational diffusion coefficient and $\eta_\text{s}$ is the viscosity of the solvent. The total force on each particle is given by $\boldsymbol{F}=\boldsymbol{F}_\text{c}+\boldsymbol{F}_\text{f}+\boldsymbol{F}_\text{r}$, where $\boldsymbol{F}_\text{c}$ represents the force due to inter-particle interactions, and and the viscous drag force is given by $\boldsymbol{F}_\text{f}=-\frac{m}{\lambda}\boldsymbol{v}$ with the damping factor defined as $\lambda=\frac{k_\text{B}T}{D}=0.12t_\text{LJ}$. The random force $\boldsymbol{F}_\text{r}$, derived from the fluctuation/dissipation theorem, has a magnitude proportional to $\sqrt{\frac{k_\text{B}Tm}{\Delta t\lambda}}$ and follows a normal distribution in both magnitude and direction. Standard Lennard-Jones units are used for time, $t_\text{LJ}=\sqrt{\sigma^2m/\epsilon_\text{LJ}}$, and energy, $\epsilon_\text{LJ}$. With this value of $\lambda$, the Schmidt number is matched between the LD simulations and the Fluid Particle Dynamics simulations described in the Supporting Information. The Schmidt number, defined as $S_c=\eta_\text{s}/\left(\rho D\right)$ with $\rho$ the fluid density, determines the crossover time between the ballistic and diffusive regimes of the particle motion.

One trajectory is collected for each system, and results are time-averaged to obtain data points at timesteps $10^i\tau_\text{B}$ with $i$ ranging from -1 to 5 in steps of 0.05. The system size is fixed with a box size of $L=34.6 \sigma$. The number of particles varies from $N_\text{p}=5930$ at the lowest considered volume fraction ($\phi=0.075$) to $N_\text{p}=15815$ at the highest considered volume fraction ($\phi=0.2$). The systems are initially equilibrated in the absence of attractive interactions, followed by the activation of attractive interactions to initiate the structure evolution over $10^5 \tau_\text{B}$. The cutoff distance for the pair potential is $5\sigma$ for the charged system and $2\sigma$ for the uncharged system. 

\subsection*{Cluster analysis}
Particles are considered part of the same cluster if $r<1.1\sigma$, roughly corresponding to the range of the attractive part of the interaction potential. Isostatic clusters have the additional criterion that each particle must have 6 or more neighbours. All clusters, including percolated clusters, are included in determining the average number of particles per cluster $\langle s\rangle$. A (isostatic) cluster is considered percolated if the maximum particle-particle distance within the cluster exceeds 0.85 times the diagonal of the simulation box. The (isostatic) percolation times in Fig.~\ref{f2} are defined as the first timestep where a percolated (isostatic) cluster is present in the system. Depercolation times are defined as the first time after the percolation time that there is no percolated cluster present in the system.

The fractal dimension $d_\text{f}$ is estimated from the radius of gyration $R_\text{g}=\left[\frac{1}{s}\sum_{i=1}^s\left(\bm{R}_i-\bm{R}_\text{c}\right)^2\right]^{1/2}$ of each cluster. Here, $\bm{R}_i$ is the center of mass of particle $i$ and $\bm{R}_\text{c}$ is the geometrical centroid of the cluster. At each timestep, the average radius of gyration $\langle R_\text{g}(s)\rangle$ as a function of cluster size $s$ is fitted with the function $\langle R_\text{g}(s)\rangle=\beta\cdot s^{d_\text{f}}$ to obtain the fractal dimension $d_\text{f}$.

\subsection*{Characteristic wavenumber analysis}
The scattering function $S(q,t)$ is determined from the 3D power spectrum of the density correlation function as $S(q,t)=\rho_q(t)\rho_{-q}(t)/N_\text{p}$. The wavenumber $q$ is inversely related to distance $r$: $q\sim 2\pi/r$. The density field is defined as $\rho(\bm{r})=\frac{6}{\pi\sigma^3}\sum_{i=1}^{N_\text{p}}\Theta\left(\frac{\sigma}{2}-| \bm{r}-\bm{R}_i | \right)$, where $\Theta$ is the Heavyside step function and $\bm{R}_i$ is the center of mass of particle $i$. The characteristic wavenumber $\langle q\rangle$ is defined as the first moment of the structure factor: $\langle q\rangle = \int qS(q)\text{d}q / \int S(q)\text{d}q$, and is inversely related to the characteristic cluster spacing or network mesh size in a cluster or gel state, respectively.

\subsection*{Rigid cluster and defect analysis}
Rigid clusters consist of tetrahedral, pyramid and pentagonal pyramid subunits formed by 4, 5 and 6 particles, respectively. Characterization of subunits is based on the topological cluster classification method described in~\cite{Royall2008,Malins2013}. Subunits that share at least 3 particles are considered to be part of the same rigid cluster. Rigidity defects are located by scanning through all pairs of rigid clusters and finding a pair of particles that meets one of the following criteria:
\begin{itemize}
    \item Particle $p_i$ in rigid cluster $i$ has the same coordinates as particle $p_j$ in rigid cluster $j$.
    \item Particle $p_i$ in rigid cluster $i$ is considered a nearest neighbour of particle $p_j$ in rigid cluster $j$. ($\triangle 1$ in Fig.~\ref{f4}c.)
    \item Particle $p_i$ in rigid cluster $i$ is connected to particle $p_j$ in rigid cluster $j$ through a cluster of particles that does not contain any rigid subunits. ($\triangle 2$ in Fig.~\ref{f4}c.)
\end{itemize}
Each time a rigidity defect is identified between particles $p_i$ and $p_j$, the nearest neighbours of particles $p_i$ and $p_j$ are excluded from the search for additional rigidity defects within the same pair of rigid clusters. The relative linearity of rigid clusters in Fig.~\ref{f6}e is averaged over particles; specifically, the value of $l_\text{rc}/l_\text{Bs}$ for a rigid cluster of size $s_\text{rc}$ is counted $s_\text{rc}$ times during the averaging process.

Bernal spirals are characterized by a linear arrangement of 4 or more face-sharing tetrahedra, where the dihedral angle $\gamma_\text{d}$ connecting the centroids of 4 neighbouring tetrahedra remains constant throughout the spiral ($2\pi/3\cdot0.9\leq \gamma_\text{d}\leq 2\pi/3\cdot1.1$ for L-spirals or $4\pi/3\cdot0.9\leq \gamma_\text{d}\leq 4\pi/3\cdot1.1$ for R-spirals). Each Bernal spiral that does not completely overlap with another larger Bernal spiral is counted as a unique Bernal spiral. When the chirality changes from L to R ($\diamond1$ in Fig.~\ref{f4}c), the L-spiral and R-spiral thus share 6 particles (3 tetrahedra). The number of chiral defects is determined by dividing rigid clusters into subclusters consisting of face-sharing tetrahedra and counting the number of times the chirality changes from L to R in each subcluster: $N_\text{L-Bs}+N_\text{R-Bs}-1$, where $N_\text{L-Bs}$ is the number of unique L-handed Bernal spirals and $N_\text{R-Bs}$ is the number of unique R-handed Bernal spirals in the subcluster.

\begin{suppinfo}
Detailed information about Fluid Particle Dynamic simulations and the effect of hydrodynamic interactions on gelation of charged colloids, an overview of the evolution of cluster size/energy distributions over time during a percolation/depercolation transition, and bond order parameter distributions of the discussed charged systems. Supporting animations illustrate the evolution of structure in the charged systems discussed in the main text.
The data that support the findings of this study are available from the corresponding author upon reasonable request.
\end{suppinfo}

\begin{acknowledgement}
This work is funded by a Rubicon grant (019.222EN.002) from the Netherlands Organization for Scientific Research (NWO). H.T. acknowledges a support by the Grant-in-Aid for Specially Promoted Research (JSPS KAKENHI Grant No. JP20H05619) from the Japan Society for the Promotion of Science (JSPS). J.O. thanks the Recommendation Program for Young Researchers and Woman Researchers of the Information Technology Center of The University of Tokyo for computational resources.
\end{acknowledgement}


\providecommand{\latin}[1]{#1}
\makeatletter
\providecommand{\doi}
  {\begingroup\let\do\@makeother\dospecials
  \catcode`\{=1 \catcode`\}=2 \doi@aux}
\providecommand{\doi@aux}[1]{\endgroup\texttt{#1}}
\makeatother
\providecommand*\mcitethebibliography{\thebibliography}
\csname @ifundefined\endcsname{endmcitethebibliography}
  {\let\endmcitethebibliography\endthebibliography}{}

\end{document}


\twocolumngrid
\section{Hydrodynamic interactions}
\label{sec_hydro}
The effect of hydrodynamic interactions on the structure evolution pathways was examined through Fluid Particle Dynamics (FPD) simulations~\cite{Tanaka2000}, which incorporate these interactions. In this method, each particle is represented as a
viscous, undeformable fluid particle expressed by $\psi_i(\bm{r})=\frac{1}{2}\left(\text{tanh}\left[\left(\frac{\sigma}{2}-\lvert \bm{r}-\bm{R}_i\rvert\right)/\xi\right]+1\right)$, where $\bm{r}$ is the position, $\bm{R}_i$ is the center of mass, and $\xi$ represents the interface thickness. This profile ensures a smooth transition between particle ($\eta_{\rm p}$) and solvent ($\eta_{\rm s}$) viscosities. For a viscosity ratio of $\eta_\text{p}/\eta_\text{s} \geq 10$, the precise value has minimal impact on the simulations, allowing particles to be effectively modeled as solid. In our simulations, $\eta_\text{p}/\eta_\text{s} = 10$ was chosen to enhance computational efficiency. 

\begin{figure*}[t!]
\includegraphics[width=.99\textwidth]{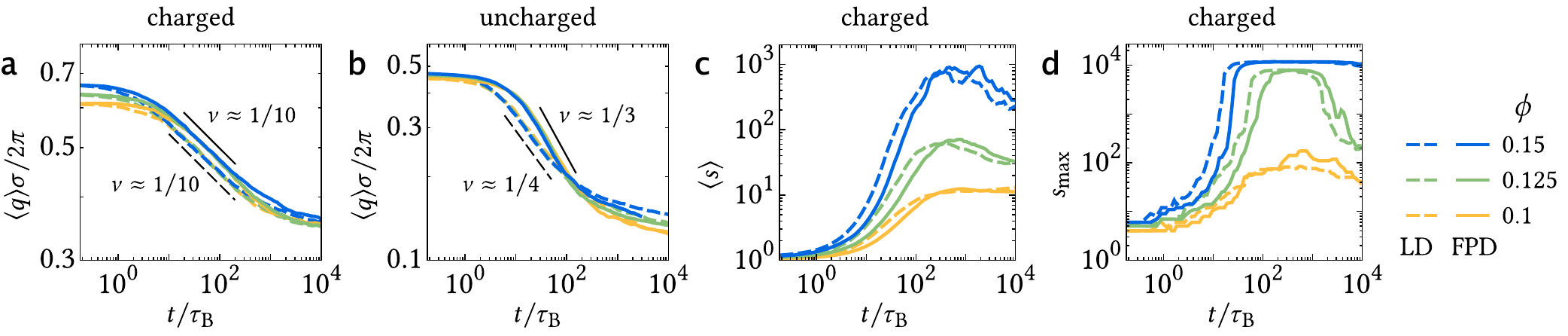}
\caption{{\bf Effect of hydrodynamic interactions on the structure evolution of charged colloidal systems.}
\textbf{a, b} Time evolution of the characteristic wavenumber, $\langle q\rangle$, for charged and uncharged systems. Solid curves represent results from FPD simulations, which include hydrodynamic interactions, while dashed curves represent results from LD simulations. Insets display the power-law scaling parameter $\nu$ ($\langle q\rangle \sim t^{-\nu}$). \textbf{c, d} Time evolution of the average cluster size, $\langle s\rangle$, and maximum cluster size, $s_\text{max}$, respectively.}
\label{Sf1}
\hypertarget{Sf1}{}
\end{figure*}

The flow field $\boldsymbol{v}$ is calculated by solving the Navier--Stokes equation
$\rho\left(\frac{\partial}{\partial t}+\boldsymbol{v} \cdot
\nabla\right) \boldsymbol{v}=\boldsymbol{f}+\nabla
\cdot\left(\boldsymbol{\sigma}+\boldsymbol{\sigma}^{\mathrm{R}}\right)$,
where $\rho$ is the constant fluid density, $\boldsymbol{f}=\sum_{i}
{\boldsymbol{F}_{i} \psi_{i}(\boldsymbol{r})}/{\int
  \psi_{i}\left(\boldsymbol{r}^{\prime}\right) \text{d}
  \boldsymbol{r}^{\prime}}$ is the grid-based force field, and
$\boldsymbol{F}_i$ is the body force of particle~$i$.
$\boldsymbol{\sigma}=\eta(\boldsymbol{r})$ $\left(\nabla
\boldsymbol{v}+(\nabla \boldsymbol{v})^{\mathsf{T}}\right)-p
\boldsymbol{I}$ is the internal stress of the fluid, where
$\eta(\boldsymbol{r})=\eta_\text{s}+\left(\eta_\text{p}-\eta_\text{s}\right)
\sum_{i=1}^{N_\text{p}} \psi_{i}(\boldsymbol{r})$ is the viscosity field,
$\boldsymbol{I}$ is the unit tensor, and $p$ is the pressure field
determined to satisfy the incompressible condition $\nabla\cdot\boldsymbol{v}=0$. The particle configurations are updated by the equation $\text{d} \boldsymbol{R}_{i}(t) / \text{d}t=\boldsymbol{V}_{i}(t)$ where $\boldsymbol{V}_{i}(t)=\int
\text{d}^3\boldsymbol{r}\left(\boldsymbol{v} \psi_{i}(\boldsymbol{r})\right)$
is the particle velocity. To incorporate thermal
fluctuations, we also introduce the random fluctuating stress field
$\boldsymbol\sigma^{\mathrm{R}}$ that satisfies the fluctuation-dissipation
relation
$\left\langle\boldsymbol{\sigma}^{\mathrm{R}}\right\rangle=\boldsymbol{0}$
and $\left\langle\sigma_{i j}^{\mathrm{R}}(\boldsymbol{r}, t)
\sigma_{k l}^{\mathrm{R}}\left(\boldsymbol{r}^{\prime},
t^{\prime}\right)\right\rangle=2 \eta(\boldsymbol{r}) k_{\mathrm{B}}
T\left(\delta_{i k} \delta_{j l}+\delta_{i l} \delta_{j k}\right)
\delta\left(t-t^{\prime}\right)$~\cite{Furukawa2018},
where $\delta$ is the Dirac delta function. We solve the NS equation by the Marker-and-Cell (MAC) method with a staggered lattice under periodic boundary condition~\cite{amsden1970simplified}.

The system was set with $L_x = L_y = L_z = 256$, a particle diameter $\sigma = 6.4$, interface thickness $\xi = 1$, thermal energy $k_\text{B}T = 1/0.07$, and time step $\Delta t = 7.3 \cdot 10^{-4}~\tau_\text{B}$. The lattice size in our system is thus $\xi$ and the unit for time is $\rho\xi^2/\eta_\text{s}$, corresponding to the time required for the ﬂuid momentum to diﬀuse over the lattice size $\xi$. The unit for energy is $\xi\eta_\text{s}^2/\rho$. The chosen parameters ensure that the Schmidt number $S_c=\eta_\text{s}/\left(\rho D\right)$, with translational diffusion coefficient $D = \frac{k_\text{B}T}{3\pi\sigma_\text{h}\eta_\text{s}\lambda^\text{T}}$, aligns with that in the LD simulations. The parameter $\lambda^\text{T} = \int \phi(\bm{r}) \text{d} \bm{r} / \int \phi(\bm{r})^2 \text{d} \bm{r}$ corrects for the smooth particle interface, and $\sigma_\text{h} = 1.04\sigma$ is used as the hydrodynamic particle diameter. Data collection and time averaging procedures mirror those in the LD simulations. Further details on the FPD simulation approach can be found in references~[\citenum{Tanaka2000,Furukawa2018, Tateno2019}].

\begin{figure*}[t!]
\includegraphics[width=.99\textwidth]{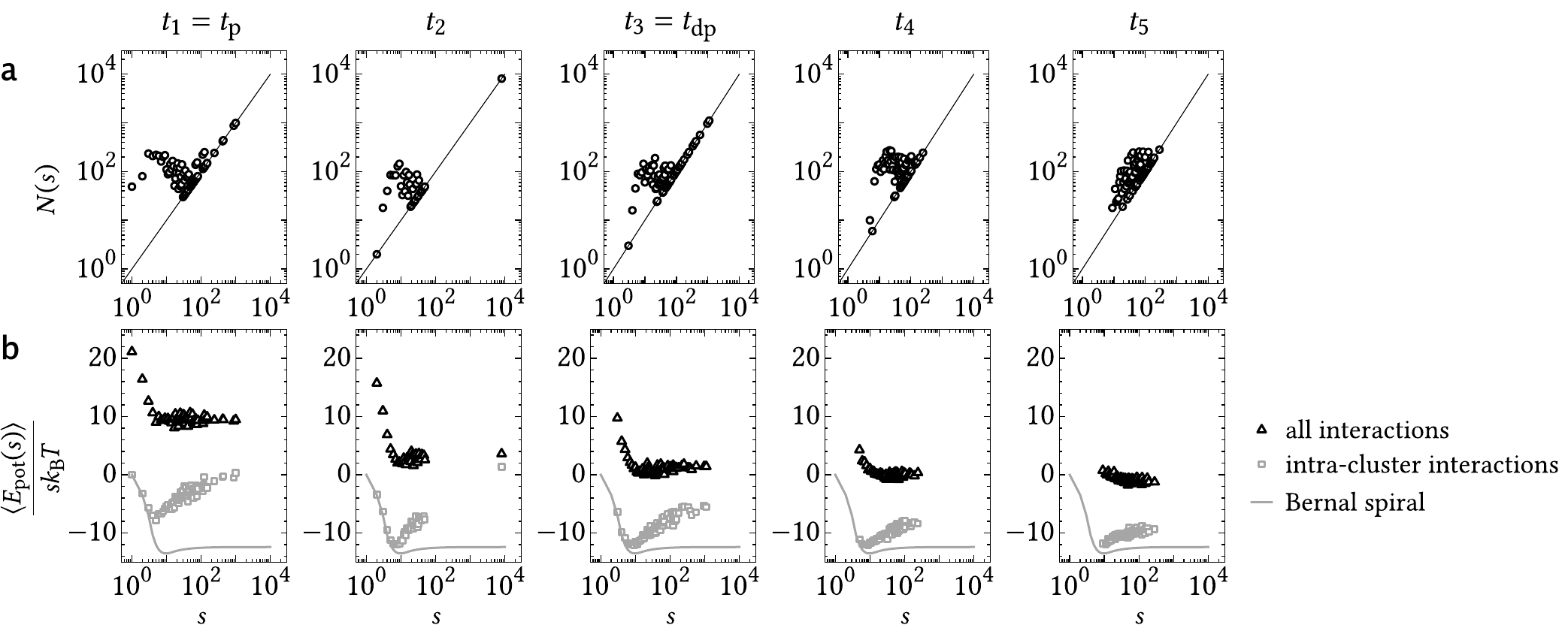}
\caption{{\bf Cluster evolution during a percolation-depercolation transition.}
\textbf{a} Total number of particles belonging to clusters with size $s$, $N(s)$, as a function of $s$ for the charged system with $\phi=0.125$. The black lines indicate $N(s)=s$, representing the presence of only one cluster with size $s$ in the system. \textbf{b} Average potential energy per particle for each cluster size $s$ as a function of $s$. Gray square symbols represent results when inter-cluster interactions are neglected, while the gray curve shows the potential energy per particle as a function of cluster size for a perfect linear Bernal spiral. The time steps are indicated as follows (the same as in the main text): percolation time ($t_\text{1}=t_\text{p}$), the time when the average cluster size $\langle s \rangle$ begins to decrease ($t_\text{2}$), depercolation time ($t_\text{3}=t_\text{dp}$), the time when the average cluster size starts increasing again ($t_\text{4}$), and the final investigated time step ($t_\text{5}=10^5\tau_\text{B}$).}
\label{Sf2}
\hypertarget{Sf2}{}
\end{figure*}

Results from the FPD simulations for systems with $\phi = 0.1$, $0.125$, and $0.15$ are presented in Fig.~\hyperlink{Sf1}{S1}. Panels a and b display the time evolution of the characteristic wavenumber $\langle q \rangle$ for charged and uncharged systems, respectively, as observed in both LD and FPD simulations. For uncharged systems, hydrodynamic interactions modify the power-law scaling behavior of network coarsening, shifting the scaling parameter $\nu$ from $1/4$ to $1/3$. The value $\nu = 1/2$, characteristic of viscoelastic phase separation~\cite{Tateno2021,Tateno2022}, is not observed due to early-stage dynamic arrest. In contrast, for charged systems, hydrodynamic interactions cause a slight delay in the onset of network coarsening, but the power-law scaling behavior remains unaffected, unlike in the uncharged systems. Hydrodynamic interactions tend to retard direct particle aggregation, which violates the incompressibility condition~\cite{Tanaka2005,Furukawa2010}. The squeezing flow of solvent resulting from approaching particles during aggregation not only delays the aggregation process, but induces the formation of elongated clusters. However, since long-range repulsive interactions inherently result in elongated aggregation, the impact of hydrodynamic interactions on particle clustering is minimal. Further analysis of the interplay between hydrodynamic and electrostatic interactions is ongoing.

Figure~\hyperlink{Sf1}{S1}c and d illustrate the time evolution of the average and maximum cluster size for charged systems, as obtained from both LD and FPD simulations. The structure evolution pathways are very similar for both simulation methods. This similarity suggests that hydrodynamic interactions are not crucial for the phenomena observed, including the percolation-depercolation transition. Therefore, the use of the less time-consuming LD simulation method in this study is justified.

\section{Cluster size and energy distributions ($\phi=0.125$)}

Figure~\hyperlink{Sf2}{S2} provides additional insights into the evolution of clusters over time for the charged system with volume fraction $\phi=0.125$, which undergoes an intermediate percolation and depercolation stage. Panel a displays the number of particles in clusters of size $s$ as a function of $s$ for the timesteps $t_1$--$t_5$ described in the caption and main text. At the timestep where the network size is largest ($t_2$), the system comprises one large percolated cluster and a distribution of smaller clusters centered around $s=10$. In subsequent stages, the network progressively breaks into smaller clusters, while the smallest clusters gradually merge into larger ones. After $t_4$, when the average cluster size begins to increase again, the distribution of particle sizes does not narrow further but slowly shifts towards larger cluster sizes.

Figure~\hyperlink{Sf2}{S2}b shows the average potential energy of particles in clusters of size $s$. Black triangles represent results including all interactions, while gray squares reflect only intra-cluster interactions. Also shown is the potential energy of a perfect Bernal spiral as a function of $s$. Over time, the potential energy for each cluster size $s$ decreases gradually due to local structural optimization. When inter-cluster repulsions are ignored, a free energy minimum appears at a cluster size of approximately $s=10$, and the potential energy distribution of clusters approaches that of perfect Bernal spirals. However, Fig.~\hyperlink{Sf2}{S2}b reveals that smaller clusters experience relatively stronger inter-cluster repulsions compared to larger clusters, and the free energy minimum is absent when all interactions are considered.

\section{Bond order parameter distributions}

Figure~\hyperlink{Sf3}{S3} displays the distributions of Steinhardt bond order parameters $q_4(i)$, $q_6(i)$, $\hat w_4(i)$, and $\hat w_6(i)$~\cite{Steinhardt1983,tenWolde1996} at $t=10^5~\tau_\text{B}$, as determined using LAMMPS. All nearest neighbors within a cutoff distance of $1.1\sigma$ were included in the computation. These parameters characterize the local order around a particle $i$ and help identify the presence of Bernal spiral structures. For a perfectly linear Bernal spiral, the bond order parameters are: $q_4=0.244$, $q_6=0.654$, $\hat w_4=0.08$, and $\hat w_6=-0.147$. Figure~\hyperlink{Sf3}{S3} demonstrates the presence of Bernal spiral structures in each system, with the characteristic peaks being most pronounced for the intermediate volume fraction $\phi=0.125$. This trend is consistent with the average size of Bernal spirals shown in Fig.~5e of the main text. While the bond order parameters indicate a significant 
\newpage

\onecolumngrid
\begin{figure*}[t!]
\includegraphics[width=.99\textwidth]{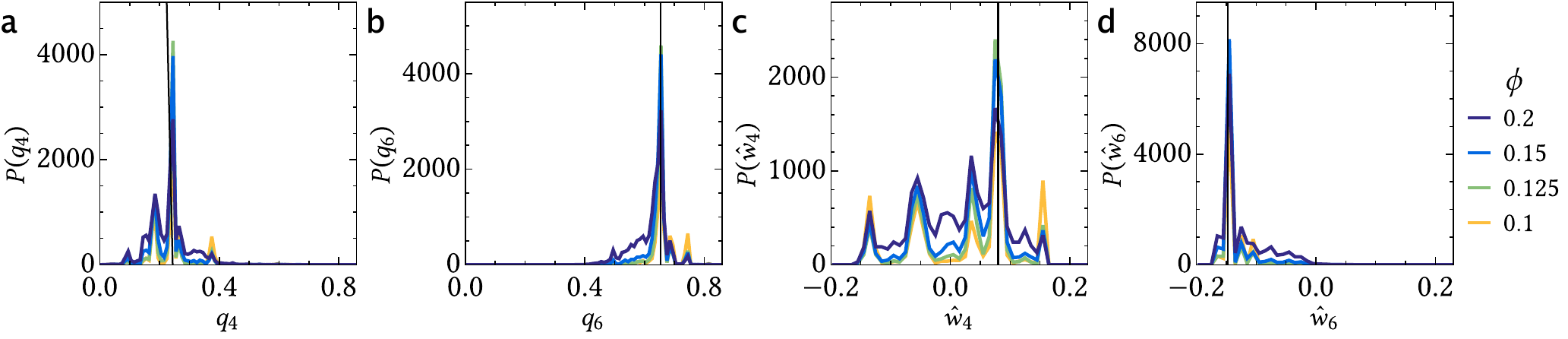}
\caption{{\bf Bond order parameter distributions.} Distributions of the bond order parameters $q_4$ (\textbf{a}), $q_6$ (\textbf{b}), $\hat w_4$ (\textbf{c}) and $\hat w_6$ (\textbf{d}) for the charged systems discussed in the main text at $t=10^5\tau_\text{B}$. Values corresponding to a perfectly linear Bernal spiral are indicated by vertical black lines.}
\label{Sf3}
\hypertarget{Sf3}{}
\end{figure*}
\twocolumngrid

\noindent presence of Bernal spiral structures, the main text details that the persistence of these structures is very short and that numerous defects cause deviations from perfect linear Bernal spirals.

\section{Supporting movies}

Supporting movies 1--4 illustrate the evolution of structure in the charged systems discussed in the main text. Particles are colored according to cluster size, with darker colors indicating larger clusters. Visualization is provided through slices that capture $0.2$ times the depth of the simulation box. Movie 1 ($\phi = 0.1$) depicts the evolution of a cluster state where percolation does not occur. Movie 2 ($\phi = 0.125$) shows the evolution of a cluster state with both percolation and depercolation transitions. Movies 3 and 4 ($\phi = 0.15$ and $\phi = 0.2$) display the evolution of a gel state, where geometric percolation is succeeded by isostatic percolation.


\providecommand{\latin}[1]{#1}
\makeatletter
\providecommand{\doi}
  {\begingroup\let\do\@makeother\dospecials
  \catcode`\{=1 \catcode`\}=2 \doi@aux}
\providecommand{\doi@aux}[1]{\endgroup\texttt{#1}}
\makeatother
\providecommand*\mcitethebibliography{\thebibliography}
\csname @ifundefined\endcsname{endmcitethebibliography}
  {\let\endmcitethebibliography\endthebibliography}{}